\documentclass{appolb}
\usepackage{graphicx}
\usepackage{amsmath} 
\usepackage{lineno}
\usepackage{url}
\usepackage{hyperref}

\begin{document}


\title{Overview of the latest ATLAS and ATLAS-AFP photoproduction results%
\thanks{Presented at ``Diffraction and Low-$x$ 2024'', Trabia (Palermo, Italy), September 8-14, 2024.
}
}
\author{Andr\'e Sopczak
\address{(on behalf of the ATLAS collaboration)\\
\vspace{2mm}
Institute of Experimental and Applied Physics,\\
Czech Technical University in Prague, and\\
Ulrich Bonse Visiting Chair for Instrumentation,\\
TU Dortmund University, Dortmund, Germany}
\address{Husova 240/5, CZ-11000 Prague}\\
\href{mailto:andre.sopczak@cern.ch}{andre.sopczak@cern.ch}
}
\maketitle
\begin{abstract}
A key focus of the physics program at the LHC is the study of head-on proton-proton collisions. However, an important class of physics can be studied for cases where the protons narrowly miss one another and remain intact. In such cases, the electromagnetic fields surrounding the protons can interact producing high-energy photon-photon collisions. Alternatively, interactions mediated by the strong force can also result in intact forward scattered protons, providing probes of quantum chromodynamics (QCD). In order to aid identification and provide unique information about these rare interactions, instrumentation to detect and measure protons scattered through very small angles is installed in the beam pipe far downstream of the interaction point. We review
photoproduction results from data collected with the ATLAS Forward Proton (AFP) and Absolute Luminosity For ATLAS (ALFA) detectors in proton-proton and heavy ion collisions.
\end{abstract}
  
\section{Introduction}
The photoproduction results are reviewed in three categories:
proton-proton interactions recorded with the ATLAS central detector~\cite{Aad:1129811},
proton-proton interactions recorded with the ATLAS-Forward-Proton (AFP) and the Absolute Luminosity For ATLAS (ALFA) detectors, and
Pb-Pb interactions recorded  with the AFP/ALFA.
The ALFA/AFP performance was recently reviewed~\cite{Sopczak:2912536}.
First, results from Pb-Pb collisions are reviewed using the central ATLAS detector and/or the AFP/ALFA detectors:
\begin{itemize}
    \item Ultra-peripheral collisions (UPC), LHC light-by-light  scattering~\cite{ATLAS:2017fur,ATLAS:2019azn},
\item Differential light-by-light and Axion-Like Particle in Pb-Pb interactions~\cite{ATLAS:2020hii},
\item
Measurements of jet production in ultra-peripheral collisions~\cite{ATLAS:2024mvt},
\item
Charged hadron yields in photonuclear collisions~\cite{ATLAS:2023pqi},
\item
Observation of the $\gamma\gamma\rightarrow\tau\tau$ process and constraints on the $\tau$-lepton anomalous magnetic moment~\cite{ATLAS:2022ryk},
\item
Search for magnetic monopoles in Pb+Pb UPC~\cite{ATLAS:2024nzp}.
\end{itemize}
Next, proton-proton collision results are reviewed using only the central ATLAS detector:
\begin{itemize}
    \item 
Electroweak, QCD and flavour physics studies, and observation of photon-induced WW production~\cite{ATLAS:2020iwi,ATLAS:2024wla}.
\end{itemize}
Subsequently, results using the central ATLAS detector and the AFP detector are reviewed:
\begin{itemize}
\item 
Observation of forward proton scattering in association with lepton pairs produced in photon fusion~\cite{ATLAS:2020mve},
\item
Di-photon resonance search with AFP tag~\cite{ATLAS:2023zfc}.
\end{itemize}

The physics program of UPC has been advancing significantly for 10 years at the LHC.
Boosted nuclei have strong electromagnetic fields and are sources of quasi-real photons
with
$E_{\rm max} < \gamma/R \approx 80$\,GeV at LHC with $Z^2(\approx  6700$) enhancement of the production cross-sections for Pb with respect to proton beam. Here, $\gamma$ is the  Lorentz factor and $R$ is the radius of the interaction.
Precision tools are needed to study the photon fluxes. An
instrumentation in the forward region is the Zero Degree Calorimeter (ZDC), which offers control over backgrounds and impact-parameter dependence.
Large integrated luminosities give access to rare processes, thus the LHC is a tool to search for physics Beyond Standard Model (BSM).

\section{Pb-Pb collisions}
\subsection{Light-by-light scattering at the LHC}

Evidence for the Standard Model (SM) process $\gamma\gamma\rightarrow \gamma\gamma$ 
was reported in lead-ion collisions in 2017~\cite{ATLAS:2017fur}. 
The observation was published subsequently~\cite{ATLAS:2019azn}.
In proton-proton collisions, the SM 
$\gamma\gamma\rightarrow \gamma\gamma$ 
has small cross-section, which can be 
enhanced in BSM, for example by an
Axion-Like-Particle (ALP).

\subsection{Differential light-by-light cross-section and Axion-Like Particle in Pb-Pb interactions}

\begin{figure}[hbp]
\vspace{-5mm}
\includegraphics[width=0.42\textwidth]{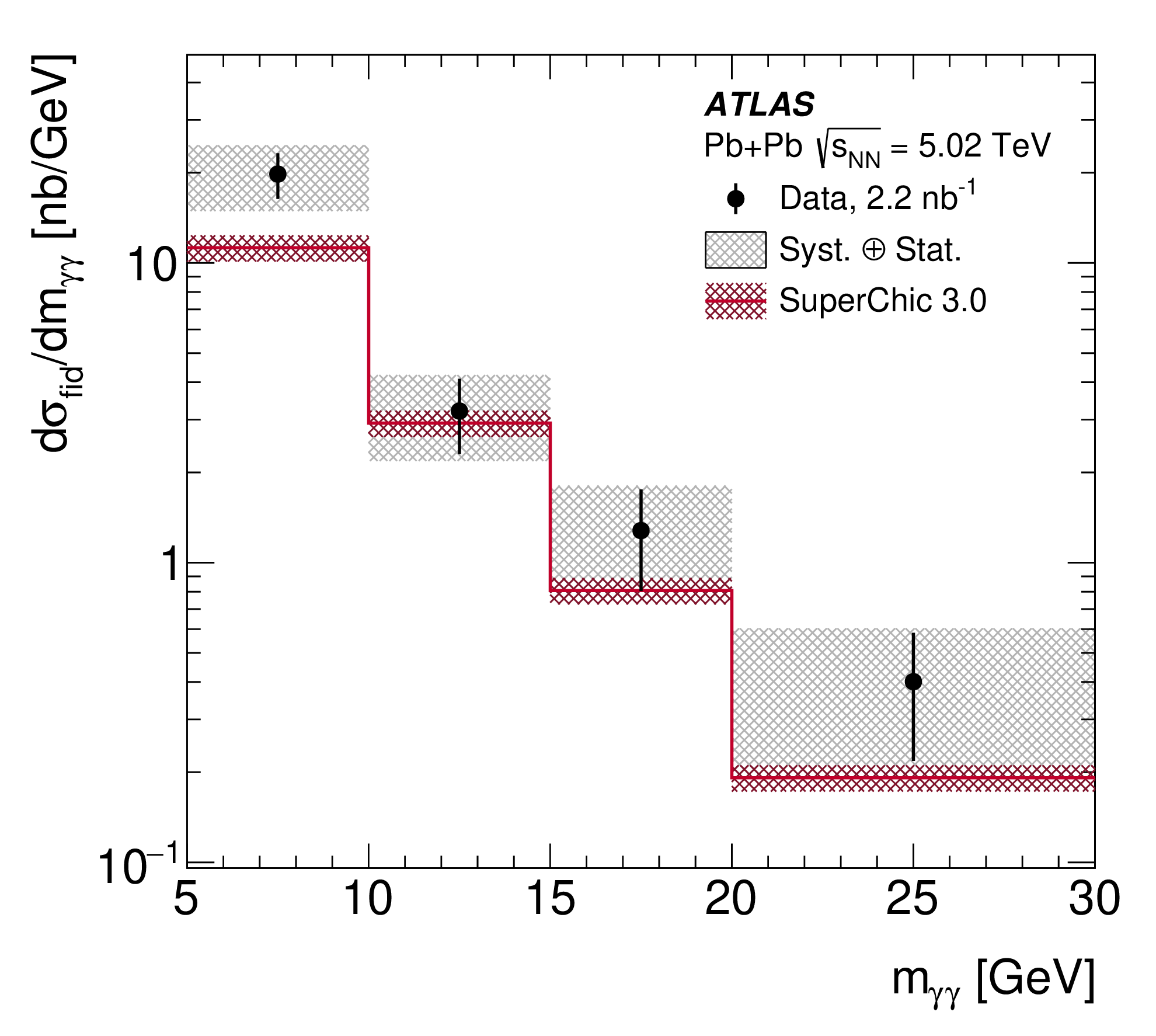}
\includegraphics[width=0.57\textwidth]{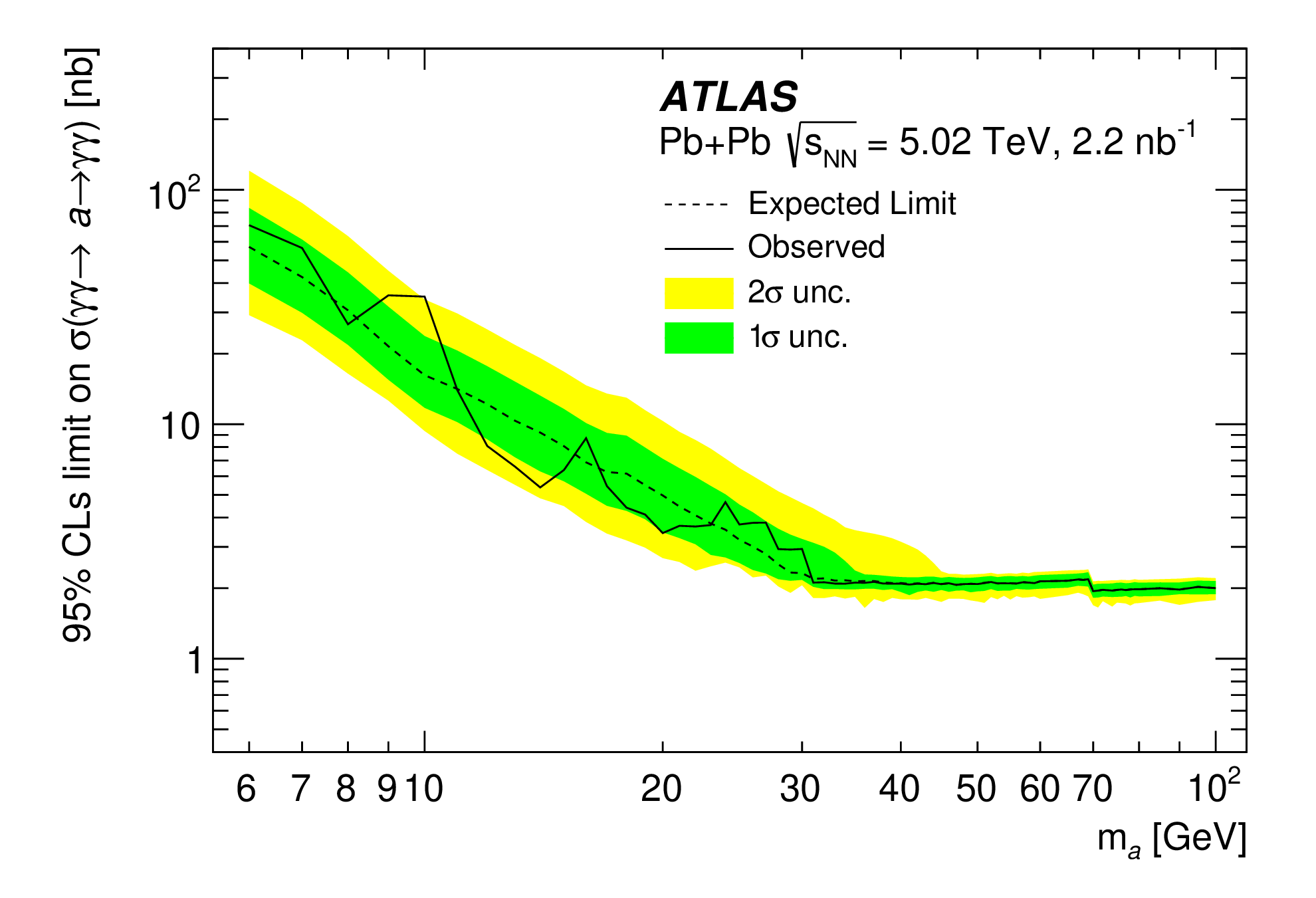}
\vspace{-6mm}
\caption{\label{fig:differential}
Left: Differential fiducial cross-sections of 
$\gamma\gamma\rightarrow \gamma\gamma$
production in Pb+Pb collisions.
The results are compared with the prediction from the SuperChic~v3.0 generator (solid line) with bands denoting the theoretical uncertainty.
Right: Upper limit on the ALP production cross-section at 95\% CL. (From Ref.~\cite{ATLAS:2020hii})
}
\vspace{-5mm}
\end{figure}

The differential light-by-light cross-section was measured in Pb-Pb collisions with
2.2\,nb$^{-1}$ of Pb+Pb data, taken 
in 2015 and 2018 at $\sqrt{s_{\text {NN}}}=  5.02$\,TeV~\cite{ATLAS:2020hii}
(Fig.~\ref{fig:differential}).
The analysis set limits at 95\% confidence level (CL) on the 
Axion-Like Particle (ALP)
production cross-section 
$\sigma(\gamma\gamma\rightarrow a \rightarrow \gamma\gamma)$
in the mass range 6 to 100\,GeV.

\subsection{Measurements of jet production in ultra-peripheral collisions}
The measurements of jet production in ultra-peripheral collisions were conducted in data sets recorded in 2018 with an integrated luminosity of 1.72\,nb$^{-1}$~\cite{ATLAS:2024mvt}.
Particle-flow jets are reconstructed using the anti-$k_t$ algorithm with radius parameter, $R=0.4$.
Triple-differential cross-sections are measured as a function of the nuclear and photon parton momentum fractions,
$x_A$ and $z_\gamma$, respectively, and the total transverse momentum in the jet system, $H_T$.
A comparison of data with the PYTHIA 8 distributions shows systematic differences which, combined with the disparity in yields may indicate contributions from processes not included in the PYTHIA 8 simulation.

\subsection{Charged hadron yields in photonuclear collisions}

\begin{figure}[hbp]
\vspace{-3mm}
\includegraphics[width=0.49\textwidth]{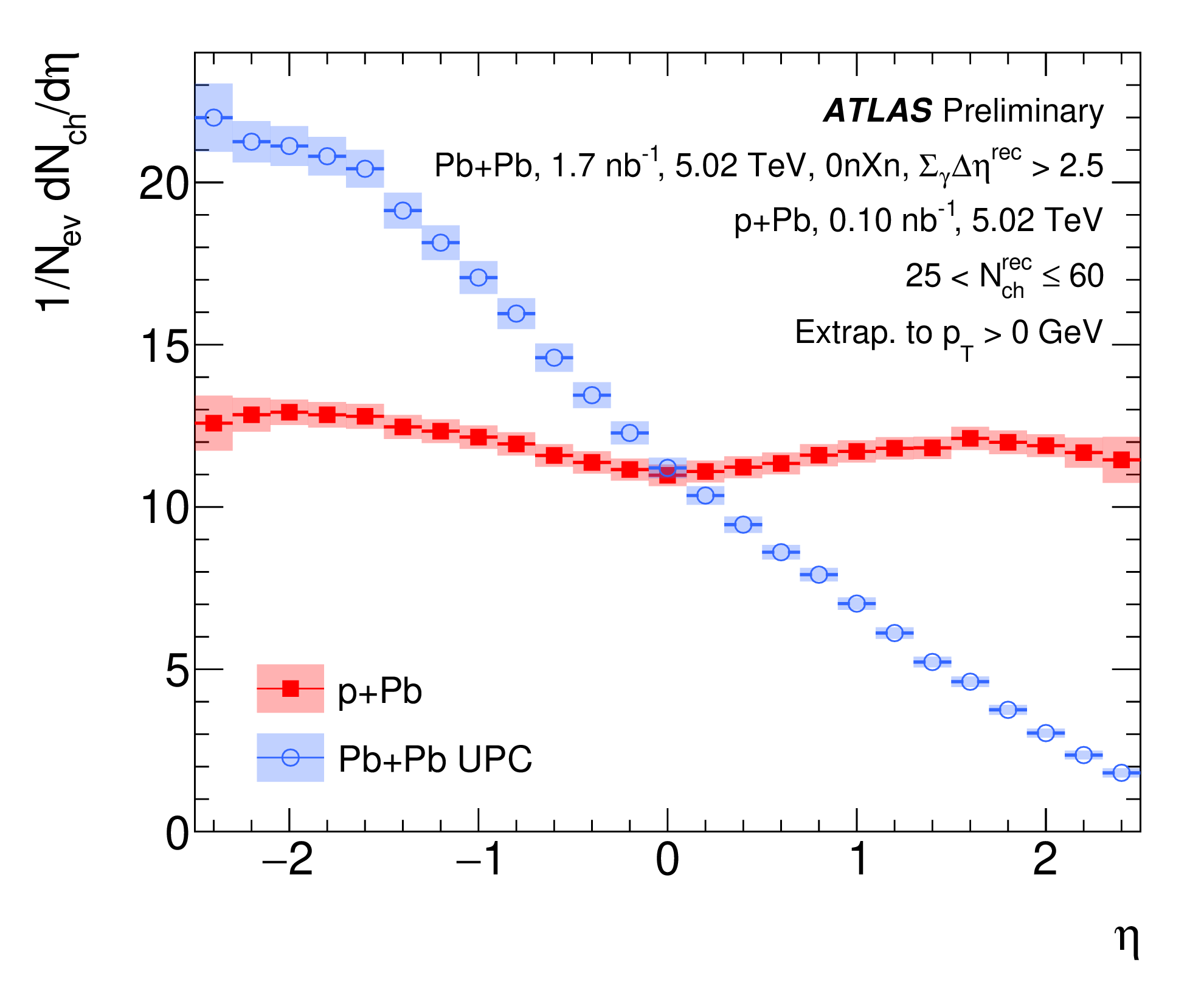}
\includegraphics[width=0.49\textwidth]{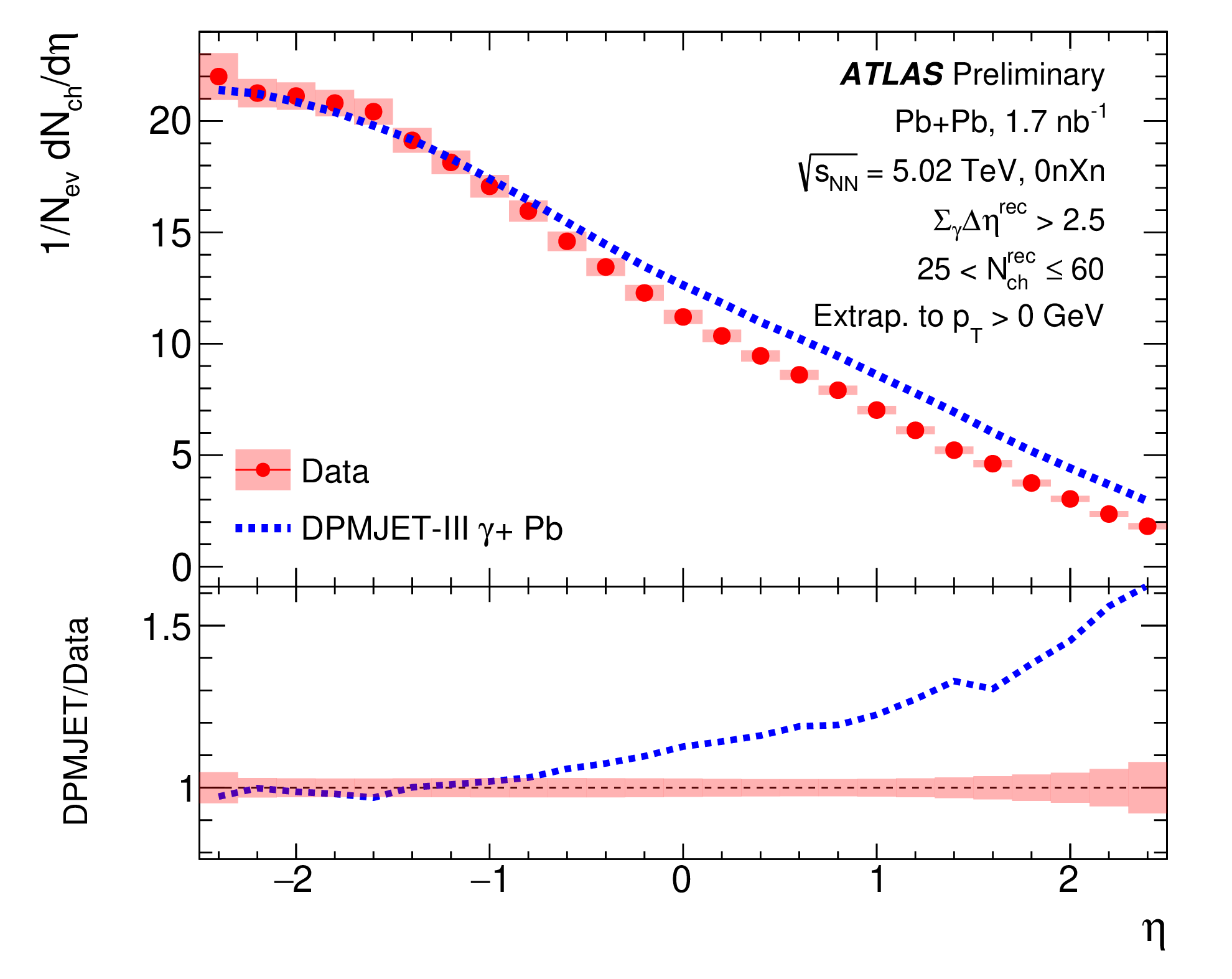}
\caption{\label{fig:hadrons}
Left: Charged-hadron yields as a function of $\eta$ for Pb+Pb UPC and p+Pb collisions.
Right: Charged-hadron yields as a function of $\eta$ in Pb+Pb UPC. Also, shown are the truth-level yield results from reconstructed event generated by DPMJET-III. (From Ref.~\cite{ATLAS:2023pqi})
}
\vspace{-3mm}
\end{figure}

Charged hadron yields in photonuclear collisions  were studied in
photo-nuclear collisions using
1.73\,nb$^{-1}$ of 5.02\,TeV Pb+Pb data collected in 2018 and
0.10\,nb$^{-1}$ of 5.02\,TeV p+Pb data collected in 2016~\cite{ATLAS:2023pqi}.
The multiplicity distribution $N$ from Pb+Pb UPC and from  p+Pb collisions are utilized in this analysis in the range $ 25 \leq N \leq 60$.
Charged-hadron yields as a function 
of $p_T$ in six $\eta$ selections are
determined from UPC Pb+Pb collisions and the p+Pb collisions (Fig.~\ref{fig:hadrons}).
Truth-level yield results from reconstructed event generated by DPMJET-III utilizing the identical
$N$ selection criteria are compared with the experimental Pb+Pb UPS data
as a function of $\eta$ (Fig.~\ref{fig:hadrons}).

\subsection{Observation of the $\gamma\gamma\rightarrow\tau\tau$ process and constraints on the 
$\tau$-lepton anomalous magnetic moment}

Photon-induced $\tau$-lepton pair production is studied in UPC Pb+Pb interactions, Pb($\gamma\gamma\rightarrow\tau\tau$)Pb, with the $\tau$-leptons decaying, for
example,  into one muon and one electron, as well as into one muon and one charged pion, and into
one muon and three charged pions~\cite{ATLAS:2022ryk}.
The muon transverse momentum distributions
are determined.

The anomalous magnetic moment, 
$a_\ell = (g  - 2)/2$ of charged leptons $\ell$ (electrons, muons, and tau) are cornerstone tests of the SM with unique sensitivity to BSM phenomena.

The $\gamma\gamma\rightarrow\tau\tau$
significance is larger than five standard deviations, and the measurement with respect to the SM expectation is
$\mu=1.03_{-0.05}^{+0.06}$ assuming the SM value for $a_\ell$.
The resulting limit
is $-0.057 < a_\ell < 0.024$ at 95\% CL.

\subsection{Search for magnetic monopoles in Pb+Pb UPC}

The Pb+Pb UPC can be used to search for magnetic monopole (MM) pair production~\cite{ATLAS:2024nzp}.
The production can occur in strong magnetic fields, primarily via the Schwinger mechanism. 
A first search in Pb+Pb interactions was reported by the MoEDAL Collaboration~\cite{MoEDAL:2021vix}.
The MM have large couplings to photons, thus perturbation theory could not be used for
the MM cross-section calculation for Heavy Ion UPC interactions. 
The production cross-section is computed non-perturbatively using semiclassical models, e.g. Free-Particle-Approximation (FPA)~\cite{Gould:2019myj}.

\begin{figure}[hbp]
\includegraphics[width=0.44\textwidth]{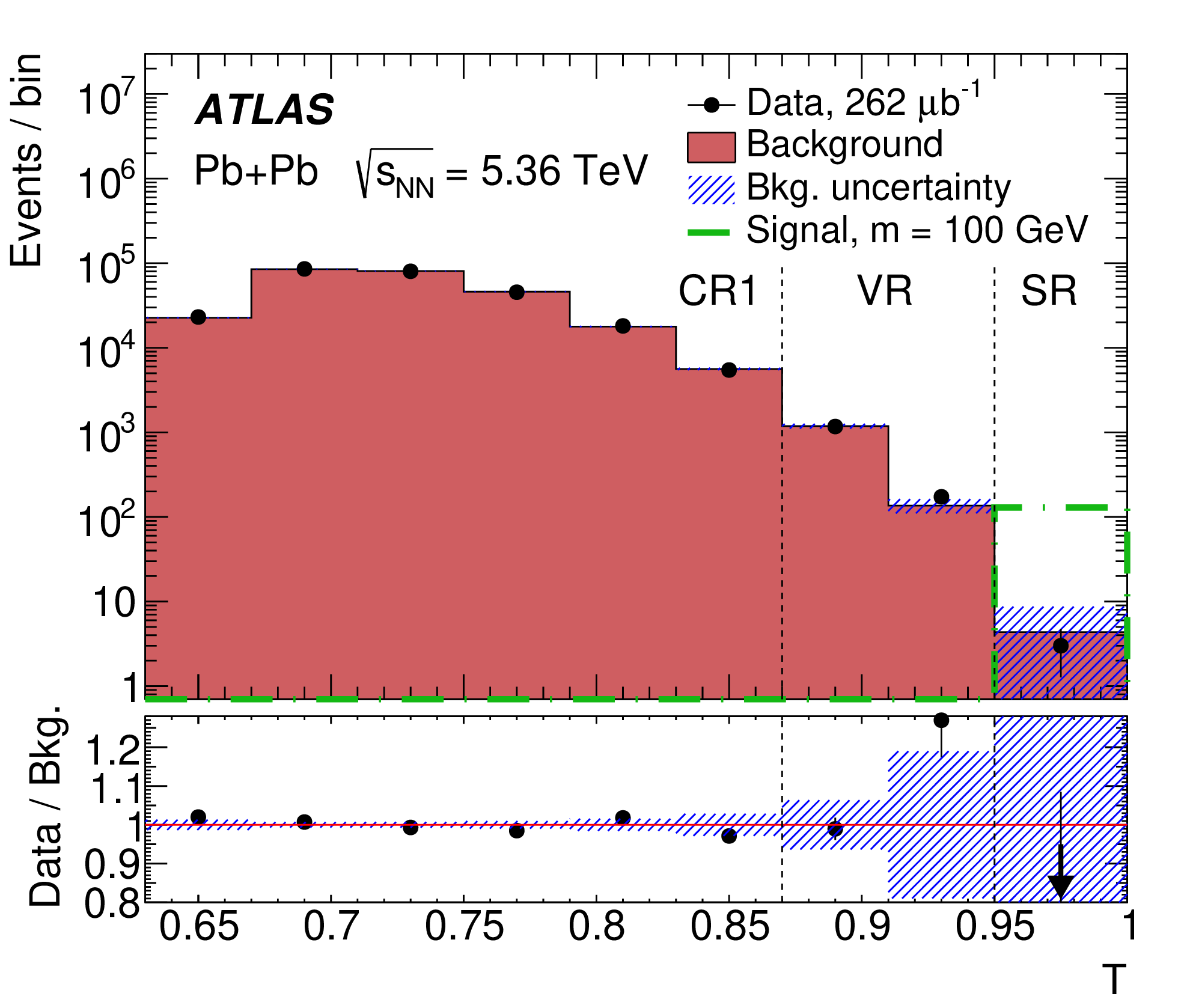}
\includegraphics[width=0.54\textwidth]{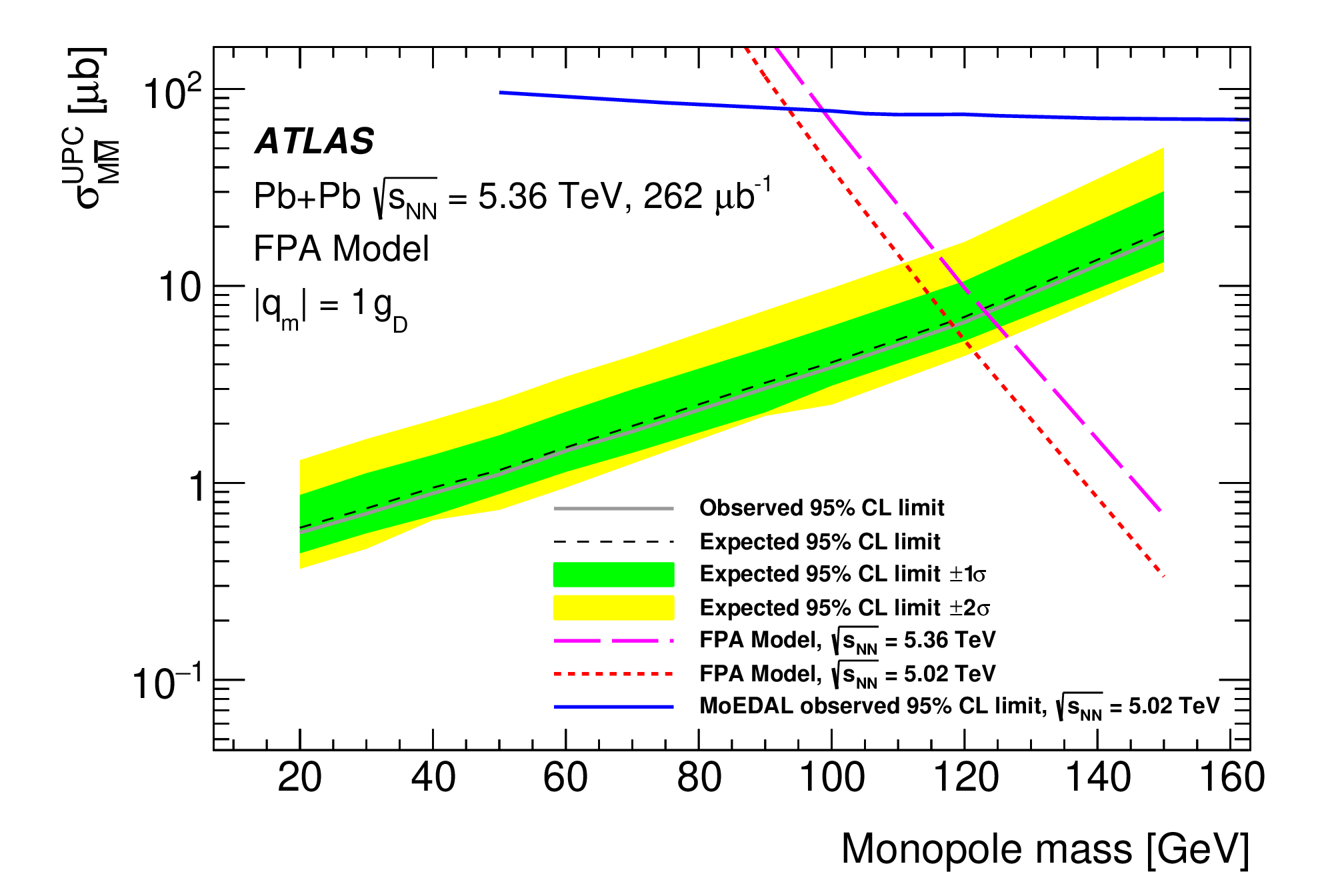}
\caption{\label{fig:mono}
Left: $T$ distribution for data in
Control Region (CR1), the 
Validation  Region (VR), and the Signal Region (SR). Data (markers) are shown together with the estimated background (filled histograms). The distributions for background use events from CR2 scaled, as described in the Ref.~\cite{ATLAS:2024nzp}. The lower panels show the ratio of data to the estimated background. The shaded bands represent the statistical uncertainty of the background. The green dotted-dashed line shows the representative signal contribution for a monopole of mass 100 GeV, and the arrow in the ratio plot is for the point that is outside the range.
Right: Expected and observed upper limits on the monopole pair-production cross-section in Pb+Pb UPC assuming the FPA model. The gray solid line (black dashed line) represents observed (expected) limits, whereas the darker and lighter shaded bands around the expected limits represent the $1\sigma$ and $2\sigma$ intervals, respectively. The limits are compared with FPA model predictions (dashed lines) and the observed limits by MoEDAL~\cite{MoEDAL:2021vix}  (blue line). (From Ref.~\cite{ATLAS:2024nzp})
}
\vspace{-6mm}
\end{figure}

Different UPC topologies are possible due to emission of neutrons.
Neutrons can be detected with the Zero Degree Calorimeters (ZDC).
LHC Run-3 Pb+Pb data of 1.6 nb$^{-1}$ at
$\sqrt{s_{\text{NN}}} = 5.36$\,TeV was used.
A selection defined the signal region with transverse thrust $T > 0.95$ (Fig.~\ref{fig:mono}).
Limits are set on the MM production in the MM mass range 20 to 150\,GeV (Fig.~\ref{fig:mono}).

\section{Proton-proton collision results using only the central ATLAS detector}

\subsection{Electroweak, QCD and flavour physics studies,  and observation of photon-induced WW production}

Proton-proton collisions at 13\,TeV 
are analysed using 139 fb$^{-1}$ Run-2 data  in di-boson system final state
$\gamma\gamma \rightarrow WW \rightarrow e\mu$~\cite{ATLAS:2020iwi,ATLAS:2024wla}.

Normalised distribution of tracks from additional proton-proton interactions, $n$, associated with the interaction vertex, in data and signal simulated with a beam spot width of $\sigma= 42$\,mm were analysed.
For data, $n$ is determined using a random $z$-position along the beam axis away from the interaction vertex.
Simulations with GEANT4 without and with beam spot width correction are used.

Some remaining background from 
$pp \rightarrow WW \rightarrow e\mu$ is taken into account, and 
the production
cross-section is  determined to be
$\sigma(\gamma\gamma\rightarrow WW)=3.12\pm0.31$ (stat.) $\pm 0.28$ (syst.)\,fb, corresponding to
a significance of 8.4 standard deviations.

\vspace{-3mm}
\section{Central ATLAS detector and the AFP detector}

\subsection{Observation of forward proton scattering in association with lepton pairs produced in photon fusion}

Scattered protons are detected by the ATLAS Forward Proton (AFP) spectrometer~\cite{ATLAS:2020mve},
and pairs of
light leptons ($ee$ or $\mu\mu$) are reconstructed in the ATLAS central detector.
The signal consists of exclusive or single dissociative (soft QCD) and re-scattering reactions.

Matching of lepton pair, $\xi_{\ell\ell}$, and proton kinematics,  $\xi_{\rm  AFP}$, is used. For the lepton kinematics $\xi_{\ell\ell}$ is determined from $m_{\ell\ell}$ and the dilepton rapidity $y_{\ell\ell}$ by momentum conservation $\xi_{\ell\ell} =(m_{\ell\ell}/\sqrt s)\exp{(\pm y_{\ell\ell})}$, where $+(-)$ corresponds to the two proton sides.
The AFP detection range is 
$0.02 < \xi_{\rm  AFP} <0.12$.
The matching requirement enhances the signal over combinatorial background ratio.

Data event candidates are di-lepton  are  presented in rapidity $y_{\ell\ell}$ versus $m_{\ell\ell}$ plane. 
Event selection and kinematic matching 
$|\xi_{\ell\ell} - \xi_{\rm  AFP}| < 0.005$ on at least one side
Shaded (hatched) areas denote the acceptance (no acceptance) for the AFP stations.
Future double tag events will increase distinction between exclusive/dissociated production.

There are 57 (123) candidates in the $ee+p$ ($\mu\mu+p$) final states.
The background-only hypothesis is rejected with a significance $>5\sigma$ in each channel.
Cross-section measurements in the fiducial detector acceptance   $\xi\in[0.035; 0.08]$ are:\\
$\sigma(ee+p) = 11.0	\pm 2.6$ (stat) $\pm1.2$ (syst) 	$\pm0.3$ (lumi)\,fb and\\
$\sigma(\mu\mu+p) = 7.2\pm 1.6$ (stat) 	$\pm 0.9$ (syst) $\pm0.2$ (lumi)\,fb.\\
A comparison with proton soft survival (no additional soft re-scattering) models agrees with the measurements.

\subsection{Di-photon resonance search with AFP tag}

In the $\gamma\gamma\rightarrow\gamma\gamma$ event, a final state proton can be intact (not dissociative)~\cite{ATLAS:2023zfc}.
Exclusive events (EL), 
single-dissociative (SD) events,
and double-dissociative (DD) event
are expected.

For the ALP production cross-section,
the coupling constant is assumed to be $f^{-1}=0.05$\,TeV$^{-1}$.
The generator SuperChic 4.02 was used for EL and  SuperChic 4.14 for SD and DD.
There are 441 events observed.
The dominant systematic uncertainty is the
AFP global alignment.

\begin{figure}[hbp]
\vspace{-2mm}
\includegraphics[width=\textwidth]{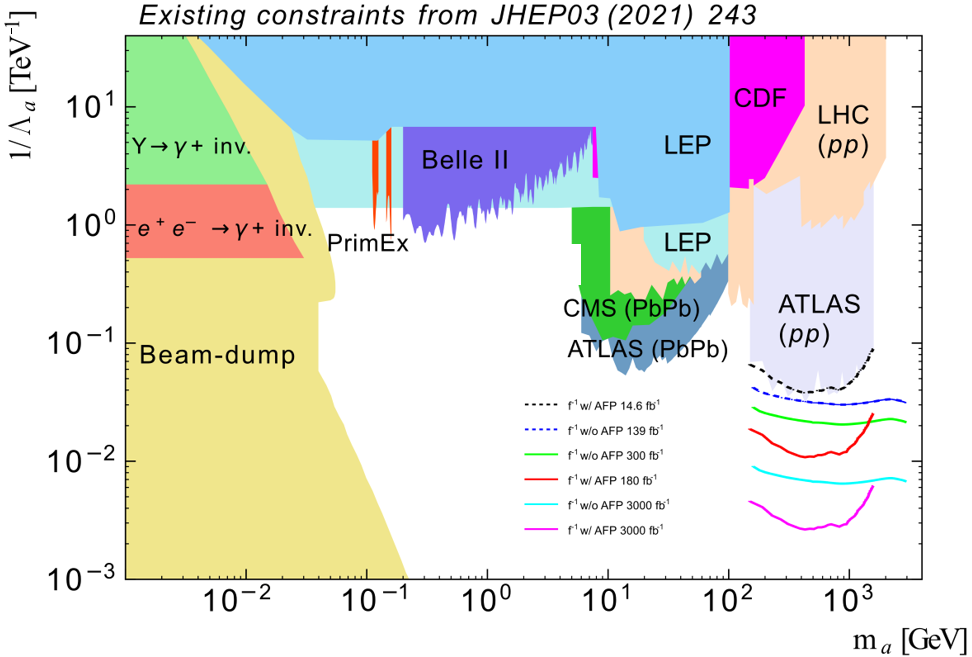}
\vspace{-5mm}
\caption{\label{fig:ALP}
ALP-coupling limit ($1/\Lambda_a = 4/f$). (From Ref.~\cite{ATLAS:2023zfc})
Also, extrapolations for Run-3 and HL-LHC are shown.}
\vspace{-5mm}
\end{figure}

Limits are set on the ALP production  
cross-section and on the production coupling constant in the ALP mass range 
200 to 1600\,GeV.
The comparison with previous $\gamma\gamma\rightarrow\gamma\gamma$
results is shown in Fig.~\ref{fig:ALP}.
The figure also shows an extrapolation (separating systematic and statistical uncertainties) for LHC Run-3 and High Luminosity (HL-LHC).

\section{Conclusions}
ATLAS has several photoproduction results based on LHC Run-2 data (taken 2015-2018) and Run-3 data (2022-) from Pb+Pb and proton-proton interactions.

ATLAS-AFP has photoproduction results from 2017 data taking in co-incidence with ATLAS central detector data
for di-lepton measurements with a proton tag, and di-photon search with a proton tag, setting limits on Axion-Like-Particles.

LHC Run-3 (2022-2026) anticipates further photoproduction results.

\section*{Acknowledgements}
The research is supported by the Ministry of
Education, Youth and Sports of the Czech
Republic under the project number LTT 17018 and and LM 2023040.
The author would like to acknowledge the DAAD support
under project number 57705645.

Copyright 2024 CERN for the benefit of the ATLAS Collaboration. CC-BY-4.0 license.
\vspace{-5mm}

\bibliographystyle{ieeetr}

\vspace*{-3mm}
\bibliography{diffraction}

\end{document}